\documentstyle[aps,epsfig,12pt]{revtex}

\def\ket#1{|#1\rangle}
\def\bra#1{\langle#1|}
\def\Lindblad#1{2a #1 a^{\dag} -a^{\dag}a #1 - #1 a^{\dag}a }

\def\tr{{\rm tr}}
\newcommand{\one}{\mbox{\tt 1}\hspace{-0.057 in}\mbox{\tt l}}

\begin{document}

\tighten

\title{Conditional evolution in single-atom cavity QED} 

\author{Andrei N. Soklakov\thanks{Direct any correspondence to:
 a.soklakov@rhul.ac.uk}\ \ and R\"udiger Schack}

%\address{Department of Mathematics, Royal Holloway, University of London, \\
%Egham, Surrey TW20 0EX, UK}

\date{25 April 2001}

\maketitle

\begin{abstract}
  We consider a typical setup of cavity QED consisting of a two-level atom
  interacting strongly with a single resonant electromagnetic field mode
  inside a cavity. The cavity is resonantly driven and the output undergoes
  continuous homodyne measurements. We derive an explicit expression for the
  state of the system conditional on a discrete photocount record. This
  expression takes a particularly simple form if the system is initially in
  the steady state. As a byproduct, we derive a general formula for the steady
  state that had been conjectured before in the strong driving limit.
\end{abstract}

\section{Introduction} \label{PhysicalSystem}

Recently there has been much experimental progress in single-atom cavity QED
\cite{HoodEtAl_2000,PinkseEtAl_2000}.  In addition to their inherent
fundamental importance, these experiments provide insight into the physics of
open quantum systems, with potential applications to, e.g., quantum
chaos~\cite{LiuEtAl_2000}, quantum control~\cite{DohertyEtAl_2000}, and
quantum computing~\cite{Steane_1998}.

\begin{figure}
\begin{center}
\epsfig{file=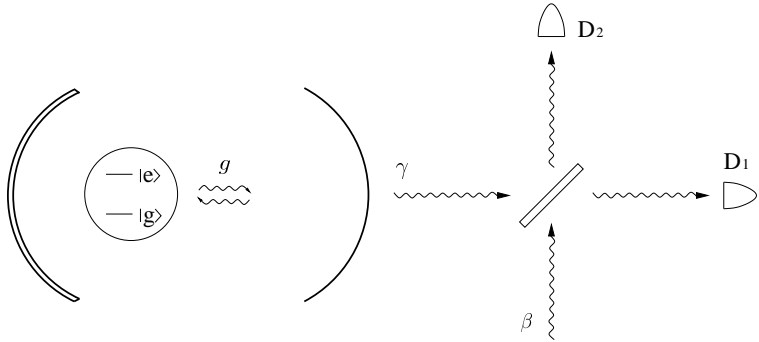,width=10cm}
\end{center}
\caption{Homodyne measurements in cavity QED. Basic parameters
of the system are the strength of the atom-cavity coupling $g$,
and the cavity field decay rate $\gamma$. The cavity is resonantly
driven by an external laser field $E$, and the cavity
output field is analyzed by the detectors $D_1$ and $D_2$
after being added to the reference field $\beta$ on the
beam-splitter.
}
\label{Cavity}
\end{figure}

In this article we consider a typical experimental setup of single-atom cavity
QED~\cite{HoodEtAl_1998}, as illustrated in Fig.~\ref{Cavity}.  The setup
consists of a single two-level atom located inside a high finesse optical
cavity, which is externally driven. A set of photodetectors is arranged to
monitor the field escaping from the system into the environment.  We assume
that the leakage of photons from the cavity mode through an output mirror is
the only significant channel through which the system interacts with the
environment. This assumption can be very realistic for high-finesse
cavities~\cite{Rempe_2000}.  Also, for simplicity, we adjust the cavity length
and the frequency of the driving field so that they both coincide with the
frequency of the atomic transition. The cavity output is monitored using
continuous homodyne measurements~\cite{PlenioKnight_1998}.  These measurements
are parameterized by one complex parameter: the reference field $\beta$ which
is added to the cavity output on a beam-splitter prior to the detection.

Given the output of the photodetectors, it is possible in principle to write
the {\it conditional\/} quantum state inside the cavity as a function of time
and the measurement record.  Usually, the conditional state is computed
numerically \cite{AlsingCarmichael_1991,MabuchiWiseman_1998} using the
formalism of stochastic master
equations~\cite{WisemanMilburn_1993b,Carmichael1993b}; these numerical
computations can require very large computational resources.  For some
experiments, however, the ability to process data in real time is
crucial~\cite{HoodEtAl_2000}. It is therefore important to develop analytical
tools for conditional state evolution.

In this paper we derive explicit expressions for the state conditioned on a
discrete homodyne measurement record in the strong coupling regime, where the
atom is strongly coupled to the intracavity field.  Our calculations are valid
for arbitrary driving field strengths. 
Our results are applicable to
experiments such as the atomic cavity
microscope~\cite{HoodEtAl_2000,PinkseEtAl_2000}, where the strong coupling is
essential, but strong driving leads to the problem of
saturation~\cite{Rempe_2000}.
We give
special attention to the experimentally important case that the system is
initially in the steady state.

The paper is organized as follows. In Sec.~\ref{MathModel} we describe the
equations that model the physical system.
In Sec.~\ref{Sec:Measurement} we review the formalism of conditional quantum
evolution and introduce the approximations for the strong coupling regime. In
Sec.~\ref{Sec:Technical} we derive a general formula for the system state
conditioned on a discrete photocount record, for an arbitrary initial system
state. In Sec.~\ref{Sec:Steady}, we give a  derivation of a general
expression for the steady state. In Sec.~\ref{Sec:Cond} we find simple
formulas for the conditional evolution in the case that the system
is initially in the steady state. We conclude in Sec.~\ref{Sec:Conclusion}.

\section{Mathematical model and main approximations} \label{MathModel}

Let $|{\rm g }\rangle$ and $|{\rm e}\rangle$ be the ground and excited states
of the atom. For simplicity, we choose the cavity length so that the
frequency of the resonant optical mode coincides with the frequency
of the atomic transition. 
Using the dipole and the rotating wave approximations
the interaction of the two-level
atom with the electromagnetic field inside the cavity
is described by the Hamiltonian~\cite{Louisell_1990}
\begin{equation}
H_{\rm int}\equiv ig(a^\dag\sigma-a\sigma^\dag)\,,
\end{equation} 
where $\sigma=|{\rm g}\rangle\langle{\rm e}|$,
$g$ is the strength of the atom-cavity coupling, and
$a$ is the annihilation operator for the intracavity field.
Including dissipation and on-resonant driving of the cavity
mode, the total unconditional master equation
in a frame rotating at the driving laser frequency reads
\begin{equation}                                             \label{OnlyStandardApprox}
\dot{\rho}=[-iH_{\rm int}
%\,\frac{g}{2}(a+a^\dag)(\sigma-\sigma^{\dag})
                  +E(a^\dag-a),\rho\,]
         +\frac{\gamma}{2}( \Lindblad{\rho} )\,,
\end{equation}
where $\rho$ is the joint density operator for the atom and the intracavity
field, $E$ is the strength of the driving, and $\gamma$ is the rate of energy
loss due to the leakage of photons from the cavity mode through an output
mirror.

From the experimental point of view the question of the steady state
is very important. In fact, using contemporary techniques it is very
difficult to prepare the system in question in any other state.
Using the Jaynes-Cummings model, Alsing and
Carmichael~\cite{AlsingCarmichael_1991} have shown
numerically that in the strong
driving limit $E\gg g$ the system approaches a steady state of the form
\begin{equation}                                                                                    \label{RhoSS1}
\rho_{\rm ss}=\frac{1}{2}( \ket{\alpha;+}\bra{\alpha;+}
                      +\ket{\alpha^*;-}\bra{\alpha^*;-})\,,
\end{equation}
where $\ket{\alpha;+}$
 and $\ket{\alpha^*;-}$ are two orthogonal
quantum states
\begin{eqnarray}                                                                                        \label{basis}
\ket{\alpha;+}&=&\frac{1}{\sqrt{2}}
               \ket{\alpha}(\ket{{\rm g}}+i\ket{{\rm e}}) 
     \equiv \ket{\alpha}\ket{+}   \,,\cr
\ket{\alpha^*;-}&=&\frac{1}{\sqrt{2}}\ket{\alpha^*}(\ket{{\rm g}}
                                 -i\ket{{\rm e}})
               \equiv \ket{\alpha}\ket{-}  \,,
\end{eqnarray}
and where $\ket{\alpha}$ is the coherent field state with amplitude
\begin{equation}
\alpha= (2E+ig)/\gamma\,.
\end{equation}
This result has been confirmed in a more recent numerical
simulation~\cite{MabuchiWiseman_1998}.
Using matrix notation for the intra-atomic degrees of freedom
in the basis $\{\ket{\pm}\}$,
Eq.~(\ref{RhoSS1}) can be rewritten in the convenient form
\begin{equation}                                                                                       \label{RhoSS}
\rho_{\rm ss}=\frac{1}{2}
                                       \left(\begin{array}{cc}
                                                 |\alpha\rangle\langle\alpha|&0\\
                                                 0& |\alpha^*\rangle\langle\alpha^*|
                                              \end{array}
                                        \right)\,,
\end{equation}
which will be useful below.
  
In this paper, we work in the strong-coupling regime ($g\gg \gamma $), which
justifies considering the evolution on time scales large compared to
$1/g$.  In Sec.~\ref{Sec:Steady} we give an analytical proof that on those
timescales, Eq.~(\ref{RhoSS1}) is a steady state of
Eq.~(\ref{OnlyStandardApprox}), for arbitrary values of the driving $E$.

\section{The measurement}  \label{Sec:Measurement}

We now rewrite
Eq.~(\ref{OnlyStandardApprox})  in the form
\begin{equation}                                                \label{MainApproximateEquation}
\dot{\rho}={\cal L}\rho\,,
\end{equation}
where the superoperator ${\cal L}$ is defined as
\begin{equation}
{\cal L}\rho\equiv
                      [-iH_{\rm int}+E(a^\dag-a),\rho\,]
                      +\frac{\gamma}{2} ( 2a\rho a^{\dag}-a^{\dag}a\rho 
                                                                -\rho a^\dag a)\,.
\end{equation}
Let the initial condition be $\rho(0)=\rho_0$.
Given superoperators ${\cal S}_0$, ${\cal J}_1$ and ${\cal J}_2$ such that
\begin{equation}
{\cal S}_0(t)=e^{({\cal L}-{\cal J}_1-{\cal J}_2)t}\,,  \label{Eq:SLJ1J2}
\end{equation}
the solution to Eq.~(\ref{MainApproximateEquation})
 can be written using a Dyson expansion,
\begin{equation}                                                                              \label{RhoDeltaT}
\rho(\Delta t)=\sum_{m=0}^{\infty}
                            \sum_{k_1,\dots,k_m} p(k_1,\dots, k_m ;\Delta t)
                                                                        \rho_{\rm c}(k_1,\dots,k_m ;\Delta t)\;,
\end{equation}
where $\tr \rho_{\rm c}(k_1,\dots,k_m ;\Delta t)=1$ and
\begin{eqnarray}                                                                                      \label{PmRm}
&& p(k_1,\dots,k_m ;\Delta t)
 \rho_{\rm c}(k_1,\dots,k_m ;\Delta t)= \cr
&& \int_0^{\Delta t}dt_{m}\cdots\int_0^{t_3}dt_2\int_0^{t_2}dt_1
{\cal S}_0(\Delta t-t_m){\cal J}_{k_m}
{\cal S}_0(t_m-t_{m-1}){\cal J}_{k_{m-1}}\cdots {\cal S}_0(t_1)\rho_0\;.
\end{eqnarray}
Following~\cite{WisemanMilburn_1993b,Carmichael1993b}
we define the ``smooth evolution'' operator
${\cal S}_0$ as
\begin{equation}                                                                                         \label{defSo}
{\cal S}_0(t)\rho\equiv N_0(t)\rho [N_0(t)]^\dag\;,
\end{equation}
where
\begin{equation}                                                                                        \label{defNo}
N_0(t)\equiv\exp\left[-iH_{\rm int}t+E(a^{\dag}-a)t
                     -\frac{\gamma}{2}(a^{\dag}a+|\beta|^2)t\right]\;,
\end{equation}
and the ``jump'' operators ${\cal J}_1$ and ${\cal J}_2$ as
\begin{equation}                                                                                            \label{Ck}
{\cal J}_{k}\rho\equiv C_k \rho C_k^{\dag}\;, {\mbox{\rm \ \ where\ \ \ }}
C_k\equiv\sqrt{\gamma/2}\,e^{i\pi(k-1)/2}[a+(-1)^k\beta]\;.
\end{equation}
The following lemma, included for completeness, shows that the definitions of
${\cal S}_0$, ${\cal J}_1$ and ${\cal J}_2$ just given are consistent with
Eq.~(\ref{Eq:SLJ1J2}).\\

\noindent{\bf Lemma 1}
{\it The above definitions satisfy the requirement
\begin{equation}                                                                               \label{SequalExp}
{\cal S}_0(t)=e^{({\cal L}-{\cal J}_1-{\cal J}_2)t}
\end{equation}
and therefore Eqns.~(\ref{RhoDeltaT}) and (\ref{PmRm}) indeed give
a solution to (\ref{MainApproximateEquation}). 
}\\
{\bf Proof}\\
Keeping  terms to  first order in $\tau$ we have
\begin{eqnarray}                                                                                    \label{SdeltaT}
  {\cal S}_0(\tau)\rho&=&N_0(\tau)\rho[ N_0(\tau)]^{\dag} \cr
 &=&\rho+\Big([-iH_{\rm int}+E(a^{\dag}-a),\rho] \cr
 &&{\phantom{=\rho+\Big(\;\;}}
                           -\frac{\gamma}{2}(a^{\dag}a\rho +\rho a^\dag a)
                                 -\gamma |\beta|^2\rho
                               \Big)\tau +O(\tau^2) \cr
 &=&(\one+\tau \cdot{\cal L})\rho
         -\gamma(a\rho a^{\dag}+|\beta|^2\rho)\tau + O(\tau^2)\,.
\end{eqnarray}
On the other hand by direct calculation we have
\begin{equation}
{\cal J}_k\rho=\frac{\gamma}{2}[a\rho a^{\dag}+(-1)^k(\beta\rho a^{\dag}
                                                    +\beta^*a\rho) +|\beta|^2\rho]
\end{equation}
which implies that
\begin{equation}
 ({\cal J}_1+{\cal J}_2)\rho=\gamma(a\rho a^{\dag}+|\beta|^2\rho)\,.
\end{equation}
Equation (\ref{SdeltaT}) therefore becomes
\begin{equation}
{\cal S}_0(\tau)\rho
= (\one+\tau\cdot [{\cal L}-({\cal J}_1+{\cal J}_2)])\rho +O(\tau^2)
\end{equation}
Taking the limit $\tau\to 0$ we have Eq.~(\ref{SequalExp}) as required.
$\Box$\\

There are many different definitions of ${\cal S}_0$, ${\cal J}_1$ and ${\cal
  J}_2$ that satisfy the above lemma. However, definitions (\ref{defSo})
and~(\ref{Ck}) are somewhat special: the quantities $\rho_{\rm
  c}(k_1,\dots,k_m ;\Delta t)$ and $p(k_1,\dots,k_m ;\Delta t)$ which they
define have an important physical meaning \cite{Carmichael1993b}.  Suppose
that the continuous measurements were performed over the time interval $\Delta
t$ and recorded as a sequence $(k_1,\dots, k_m ;\Delta t)$ of photodetector
labels in the order of photodetections. For example, $k_j=1$ would mean that
the $j$th photodetection was registered by the first detector.  Then the
probability of the measurement record $(k_1,\dots, k_m ;\Delta t)$ is given by
$p(k_1,\dots, k_m ; \Delta t)$, and the corresponding conditional state is
$\rho_{\rm c}(k_1,\dots,k_m ;\Delta t)$.

We will now prepare to consider the conditional system evolution
on time scales large compared to $1/g$.
First we notice that
\begin{equation}
H_{\rm int}=H_0+H_1\,,
\end{equation}
where
\begin{equation}
\begin{array}{llll}
H_0&\equiv& -g(a^\dag+a)\sigma_y/2\,, &
                           \ \ \ \sigma_y\equiv i(\sigma^\dag-\sigma) \cr
H_1&\equiv& ig(a^\dag-a)\sigma_x/2\,, &
                           \ \ \ \sigma_x\equiv \sigma^\dag+\sigma\,.
\end{array}
\end{equation}
We define
\begin{equation} \label{defQ}
Q\equiv \exp(-iH_0t-iH_1t+Ft)\,,
\end{equation}
where
\begin{equation}
F\equiv E(a^\dag-a)-\gamma a^\dag a/2\,.
\end{equation}
These definitions are connected to the definition~(\ref{defSo}) of the
smooth evolution operator via the relation
\begin{equation}
N_0\equiv e^{-\gamma|\beta|^2t/2}Q\,.
\end{equation}
We rewrite $Q$ in the form
\begin{equation}                              \label{QRo}
Q=e^{-iH_0t}R_0\,,
\end{equation}
so that
\begin{equation}
\frac{dQ}{dt}=-iH_0e^{-iH_0t}R_0+e^{-iH_0t}\frac{dR_0}{dt} {\,.}
\end{equation}
From the definition~(\ref{defQ})
and Eq.~(\ref{QRo}) we have:
\begin{equation}
\frac{dQ}{dt}=(F-iH_0-iH_1)e^{-iH_0t}R_0\,.
\end{equation}
Combining the last two equations we obtain that $R_0$ obeys
the equation
\begin{equation}                                                        \label{dRoOVERdt}
\frac{dR_0}{dt}=\left(X(t)+e^{iH_0t}Fe^{-iH_0t}\right)R_0\,,
\end{equation} 
where
\begin{equation}
X(t)\equiv - e^{iH_0t} iH_1 e^{-iH_0t}\,.
\end{equation}
Using the Corollary to Theorem 1 from the Appendix together with
the identity $e^{-ya^\dag}a=(a+y) e^{-ya^\dag}$ we obtain
\begin{eqnarray}
2X(t)&=&g\,e^{iH_0t}(a^\dag-a)\sigma_x e^{-iH_0t}\cr
       &=&g\,(a^\dag-a-ig\sigma_yt)e^{iH_0t}\sigma_x e^{-iH_0t}\,.
\end{eqnarray}
The identity $e^{iA\sigma_y}=\cos A+i\sigma_y\sin A$ gives
\begin{equation}
e^{iH_0t}\sigma_x e^{-iH_0t}
 =\sigma_x\cos[gt(a^\dag+a)]-\sigma_z\sin[gt(a^\dag+a)]\,,
\end{equation}
where $\sigma_z\equiv i\sigma_y\sigma_x$.
Finally we obtain
\begin{equation} 
2X(t)=g(a^\dag -a -ig\sigma_y t)
\left( \sigma_x\cos[gt(a^\dag+a)] -\sigma_z\sin[gt(a^\dag+a)]
\right)\,.
\end{equation}
At time scales large compared to $1/g$ we can neglect terms oscillating at
frequency $1/g$ in Eq.~(\ref{dRoOVERdt}). This means we can set
$X(t)=0$ in Eq.~(\ref{dRoOVERdt}), which becomes
\begin{equation}                                     \label{APPROXdRoOVERdt}
\frac{dR_0}{dt}\approx e^{iH_0t}Fe^{-iH_0t} R_0\,.
\end{equation} 
This approximation has some similarity with the standard  rotating-wave
approximation. 

Now consider the operator
\begin{equation}
M\equiv \exp (-iH_0t+Ft)\,.
\end{equation}
Using the same technique as in Eqns.~(\ref{QRo} -- \ref{dRoOVERdt})
it is easy to show that
\begin{eqnarray}
M=e^{-iH_0t} R_1\,,
\end{eqnarray}
where $R_1$ obeys the equation:
\begin{equation}
\frac{dR_1}{dt}= e^{iH_0t}Fe^{-iH_0t} R_1\,.
\end{equation}
This equation coincides with (\ref{APPROXdRoOVERdt}), which means
that at time scales $\delta t\gg 1/g$ the operator $Q$ can be replaced
with $M$. The smooth evolution ${\cal S}_0$ can therefore be
 approximated as
\begin{equation}                                                                                     \label{SoS}
{\cal S}_0\approx {\cal S} \,,
\end{equation}
where ${\cal S}$ is defined as
\begin{equation}                                                                                    \label{defS}
{\cal S}(t)\rho\equiv N(t)\rho [N(t)]^\dag\,,
\end{equation}
and where
\begin{equation}                                                                                     \label{defN}
N(t)\equiv\exp\left[-iH_0t+E(a^{\dag}-a)t
                     -\frac{\gamma}{2}(a^{\dag}a+|\beta|^2)t\right]\,.
\end{equation}

\section{Conditional evolution for arbitrary initial states}   \label{Sec:Technical}

In this section we derive a general formula for the state, $\rho_{\rm
  c}(k_1,\dots, k_m ;\Delta t)$, conditioned on a discrete photocount record
for an arbitrary initial state. The formula is a direct consequence of two
technical theorems, whose proofs are given in the Appendix. 
At time scales $\delta t\gg 1/g$, the theorems allow us to simplify
Eq.~(\ref{PmRm}) by changing the order in
which the smooth evolution operators ${\cal S}$ and the jump operators
${\cal J}_k$ appear.

Using these theorems we can proceed with the calculation of the
conditional density matrix $\rho_{\rm c}(k_1,\dots, k_m ;\Delta t)$.
We have from Eqns.~(\ref{PmRm}), (\ref{Ck}) and
Eqns.~(\ref{SoS}--\ref{defN}) that
\begin{eqnarray}
&& \hspace{-10mm} p(k_1,\dots, k_m ;\Delta t)\rho_{\rm c}(k_1,\dots, k_m ;\Delta t)\cr
&&  \hspace{-8mm} \approx \frac{1}{m!} \int_0^{\Delta t}dt_m\cdots
        \int_0^{\Delta t}dt_{2}\int_0^{\Delta t}dt_1
[N(\Delta t-t_m)C_{k_m}\cdots N(t_2-t_{1})C_{k_{1}}N(t_1)]
                                         \rho_0[\cdots]^{\dag}\;.
\end{eqnarray}
We can now use Theorem~2 to compute the operator in the square
brackets. We have, for instance,
\begin{equation}
C_{k_{1}}N(t_1)=N(t_1) f_k\bigg[e^{-\gamma t_1/2}a
                                    +\frac{1-e^{-\gamma t_1/2}}{\gamma}
                                       (2E+ig\sigma_y)+(-1)^{k_1}\beta\bigg]\,.
\end{equation}

Then, using the identity $N(t_2-t_1)N(t_1)=N(t_2)$, we see that repeating
the same type of calculations we have
\begin{eqnarray}
&& N(\Delta t-t_m)C_{k_m}\cdots N(t_2-t_{1})C_{k_{1}}N(t_1)=\cr
&&\hspace{15mm} N(\Delta t)\prod_{p=1}^{m}f_{k_p} \bigg[e^{-\gamma t_p/2}a
                                    +\frac{1-e^{-\gamma t_p/2}}{\gamma}
                                       (2E+ig\sigma_y)+(-1)^{k_p}\beta\bigg]\,.
\end{eqnarray}
Using the identity $f_kf_k^*=\gamma/2$ we therefore have
\begin{equation}                                                                                 \label{pMrhoM}
p(k_1,\dots, k_m ;\Delta t)\rho_{\rm c}(k_1,\dots, k_m ;\Delta t)
                                            = \frac{\gamma^m}{2^mm!}
                                            N(\Delta t)G(\rho_0,\beta)N^{\dag}(\Delta t)\,,
\end{equation}
where
\begin{eqnarray}                                                                               \label{RrhoSub0}
G(\rho_0,\beta) 
&=&
\int_0^{\Delta t}dt_m\int_0^{\Delta t}dt_{m-1}\cdots\int_0^{\Delta t}dt_1 \cr
&& \bigg(\prod_{p=1}^{m}\bigg[e^{-\gamma t_p/2}a
                                    +\frac{1-e^{-\gamma t_p/2}}{\gamma}
                                       (2E+ig\sigma_y)+(-1)^{k_p}\beta \bigg]\bigg)
                                         \rho_0\bigg(\cdots\bigg)^{\dag}\;.
\end{eqnarray}
For notational convenience, we do not indicate explicitely the dependence
of $G(\rho_0,\beta)$
on the measurement record $(k_1,\dots, k_m ;\Delta t)$
which, however, should always be remembered.

Equations (\ref{pMrhoM}) and~(\ref{RrhoSub0}) have a relatively simple
structure. The terms $N(\Delta t)$, which are given in factored form by
Theorem 1, are the same for all possible measurement records. This means
that all the information about the measurement records is contained in 
the function $G(\rho_0,\beta)$. 
The integrand in $G(\rho_0,\beta)$ is a polynomial in $a$, $\sigma_y$ and
$\rho_0$. The scalar coefficients of this polynomial are constants or
proportional to either $e^{-\gamma t_p/2}$ or $e^{-\gamma t_p}$. Therefore all
the integrals in Eq.~(\ref{RrhoSub0}) can be easily evaluated, so that
$G(\rho_0,\beta)$ takes the form of a polynomial in $a$, $\sigma_y$ and
$\rho_0$ with known coefficients. In this way, Eqns.~(\ref{pMrhoM})
and~(\ref{RrhoSub0}) provide an explicit solution for the conditional
evolution on the time scales considered.

\section{Derivation of the steady state}   \label{Sec:Steady}

In this section we show that, at timescales $\delta t\gg 1/g$,
 the state $\rho_{\rm ss}$ defined
by~(\ref{RhoSS}) is a steady state of
the master equation~(\ref{MainApproximateEquation}).
Notice that the only free parameter in our homodyne measurements is
the complex parameter $\beta$. If we can find a value of $\beta$ 
such that for any measurement record $(k_1,\dots, k_m ;\Delta t)$,
the conditional density matrix satisfies
\begin{equation}                     \label{BlankMeasurement1}
  \rho_{\rm c}(k_1,\dots, k_m ;\Delta t) =\rho_{\rm ss}\;,
\end{equation}
then $\rho_{\rm ss}$ must be a steady state.
This is because the solution~(\ref{RhoDeltaT}) of the
unconditional master equation~(\ref{MainApproximateEquation}) becomes, 
in this case,
\begin{equation}                         \label{BlankMeasurement2}
  \rho(\Delta t)=\sum_{m=0}^{\infty}\sum_{k_1,\dots, k_m}
p(k_1,\dots, k_m ;\Delta t)\rho_{{\rm ss}}=\rho_{\rm ss}
\end{equation}
for any $\Delta t$.  Intuitively, one would expect that, if subjected to a
nontrivial measurement, the system would normally depart from the steady
state. In our case, however, we will find that Eq.~(\ref{BlankMeasurement1}) is
satisfied for all real values of $\beta$.

Before we proceed with our rigorous analysis it may be helpful to
develop some intuition about the dependence of the conditional
evolution on $\beta$. In particular we are interested in the dependence
of the conditional evolution on the phase $\phi=\arg\beta$. 
In our analysis we deal with the conditional evolution conditioned
on a discrete photocount record, which is the most general case.
However, a lot of insight about the dependence of the conditional
evolution on the phase $\phi$ can be gained by taking the limit
$|\beta|\to\infty$. In this limit the detectors are registering
continuous photocurrents rather than discrete photocounts. 
Because the resulting measurement records can be viewed
as continuous functions of time it becomes possible to derive
a master equation for the conditional density matrix~$\rho_{\rm c}$.
According to Ref.~\cite{WisemanMilburn_1993b}, this can be done by
taking the double limit $|\beta|\propto\epsilon^{-1}\to\infty$
and $\gamma \Delta t\propto \epsilon^{3/2}\to 0$ in the Dyson
expansion (\ref{RhoDeltaT}).
If the measurement record consists of the difference photocurrent
$I_{-}=I_2-I_1$, where $I_1$ and $I_2$ are the photocurrents
detected by the first and the second detectors respectively, then
the resulting master equation for the conditional density
matrix becomes~\cite{WisemanMilburn_1993a,WisemanMilburn_1993b}
\begin{equation}                                                                          \label{StochasticME}
\dot{\rho}_{\rm c}=
                  {\cal L}\rho
           +\sqrt{\gamma\eta}
              \Big(e^{-i\phi}a\rho_{\rm c}+e^{i\phi} \rho_{\rm c}a^{\dag}
                 -\tr\!\left[\rho_{\rm c}(e^{-i\phi}a+e^{i\phi}a^{\dag})\right]
           \rho_{\rm c}
              \Big)\xi\,,
\end{equation} 
where $\eta$ is the efficiency of the photodetection,
and $\xi$ is the Gaussian white noise which,
in practice, should be taken from experimental observations of the
difference photocurrent $I_{-}$ 
via the relation
\begin{equation}                                                                   \label{DiffPhotocurrent}
I_{-}
=|\beta|\left(\gamma\eta\;\tr\!\left[\rho_{\rm c}(e^{i\phi}a^\dag+e^{-i\phi} a)
                                               \right]
                  +\sqrt{\gamma\eta}\xi\right)\;.
\end{equation}
Compared to the unconditional master
equation~(\ref{OnlyStandardApprox}),
Eq.~(\ref{StochasticME}) has an additional term
\begin{equation}
  \sqrt{\gamma\eta}
\Big(e^{-i\phi}a\rho_{\rm c}+e^{i\phi}\rho_{\rm c} a^\dag
                             -\tr\!\left[\rho_{\rm c}(e^{-i\phi}a+e^{i\phi}a^\dag)\right]
                  \rho_{\rm c}\Big)\xi\,,
\end{equation} 
which, for $\phi=0$ and $\rho_{\rm c}=\rho_{\rm ss}$, is proportional to
$\rho_{\rm ss}$. This means that, if ${\cal L}\rho_{\rm ss}=0$, i.e., if
$\rho_{\rm ss}$ is a steady state of the unconditional evolution, then
conditional and unconditional evolution coincide for $\phi=0$.  
This situation is similar to the one described by
Eqns.~(\ref{BlankMeasurement1}) and (\ref{BlankMeasurement2}), which suggests
to consider the case of real $\beta$ in the following rigorous derivation.

We now substitute $\rho_0=\rho_{\rm ss}$ from Eq.~(\ref{RhoSS}) into
Eq.~(\ref{RrhoSub0}), keeping $\beta$ arbitrary for the moment.
We obtain
\begin{equation}                                                               \label{NewEquationForR}
G(\rho_{{\rm ss}},\beta) =
\int_0^{\Delta t}dt_m\int_0^{\Delta t}dt_{m-1}\cdots\int_0^{\Delta t}dt_1
       \Big( \prod_{p=1}^{m}e^{-\gamma t_p/2}[a + f(\sigma_y,\beta)]
        \Big)
                 \rho_{{\rm ss}}
       \Big( \cdots \Big)^{\dag}\;,
\end{equation}
where
\begin{equation}
f(\sigma_y,\beta)\equiv\frac{e^{\gamma t_p/2}-1}{\gamma}
                                       (2E+ig\sigma_y)+(-1)^{k_p}\beta e^{\gamma t_p/2}\,.
\end{equation}
We note that
\begin{equation}
[\sigma_y, \rho_{\rm ss}]=0\ \ \ {\rm and}\ \ \ (\sigma_y)^2=\one\,.
\end{equation}
Using the first of these properties and the expression for $\rho_{\rm ss}$
as given by Eq.~(\ref{RhoSS}), we have by direct calculation
\begin{eqnarray}                                                                                      \label{aPlusF}
[a+f(\sigma_y,\beta)]\rho_{\rm ss} [a+f(\sigma_y,\beta)]^{\dag}
&=& \Big( f(\sigma_y,\beta) [f(\sigma_y,\beta)]^{\dag}
          +2{\rm Re}[f(\sigma_y,\beta)] {\rm Re}(\alpha) \cr
&&    +2{\rm Im}[f(\sigma_y,\beta)] {\rm Im}(\alpha)\sigma_y
           +|\alpha|^2
\Big)\rho_{\rm ss}\,,
\end{eqnarray}
where $\alpha=(2E+ig)/\gamma$. We will use this equation for imaginary $\beta$
in the next section.

For the rest of this section, we assume that $\beta$ is real. Using this 
and the fact that $(\sigma_y)^2=\one$, we find that 
\begin{equation}
f(\sigma_y)[f(\sigma_y)]^{\dag}
=\frac{4E^2+g^2}{\gamma^2}[e^{\gamma t_p/2}-1]^2
   +\beta^2e^{\gamma t_p}
   +\frac{4E\beta}{\gamma}(-1)^{k_p}
      (e^{\gamma t_p}-e^{\gamma t_p/2})\,,
\end{equation}  
and
\begin{equation}
{\rm Re}[f(\sigma_y)] {\rm Re}(\alpha) 
+{\rm Im}[f(\sigma_y)] {\rm Im}(\alpha)\sigma_y
=\frac{4E^2+g^2}{\gamma^2}(e^{\gamma t_p/2}-1)
   +\frac{2E\beta}{\gamma}(-1)^{k_p}e^{\gamma t_p/2}\,.
\end{equation}
Because $|\alpha|^2=|2E+ig|^2/\gamma^2=(4E^2+g^2)/\gamma^2$
we therefore have according to Eq.~(\ref{aPlusF}):
\begin{equation}
  [a+f(\sigma_y,\beta)]\rho_{{\rm ss}}[a+f(\sigma_y,\beta)]^\dag
 = e^{\gamma t_p} [\frac{4E^2+g^2}{\gamma^2}
                                     +(-1)^{k_p}\frac{4E\beta}{\gamma}+\beta^2]\,.
\end{equation}
Substituting this into (\ref{NewEquationForR}) we obtain:
\begin{equation}                                                                              \label{RrhoSubSS}
G(\rho_{\rm ss},\beta) =(\Delta t)^m \prod_{p=1}^{m}
\Big(
\frac{4E^2+g^2}{\gamma^2}+(-1)^{k_p}\frac{4E\beta}{\gamma}+\beta^2
\Big)
      \rho_{\rm ss}\;.
\end{equation}
Therefore, according to Eq.~(\ref{pMrhoM}),
\begin{equation}                                                     \label{IDontKnowHowToCallIt}
\rho_{\rm c}(k_1,\dots,k_m ; \Delta t)
\propto N(\Delta t)\rho_{\rm ss}N^{\dag}(\Delta t)\,,\ \ 
\ {\rm  for\ any\ real}\ \beta\,.
\end{equation}
As the final step of our argument, we now prove a lemma that, together with
Eq.~(\ref{IDontKnowHowToCallIt}) and
the normalization of the density matrix, implies Eq.~(\ref{BlankMeasurement1}).\\

\noindent{\bf Lemma 2} {\it
Smooth evolution leaves $\rho_{\rm ss}$ invariant in the following sense:}
\begin{equation}
 N(\Delta t)\rho_{\rm ss}N^{\dag}(\Delta t)\propto \rho_{\rm ss}\,.
\end{equation}
{\bf Proof}\\
Because $\sigma_y$ and  $\rho_{\rm ss}$ commute, we can see from
Eq.~(\ref{RhoSS}) that the smooth evolution leaves $\rho_{\rm ss}$ diagonal:
\begin{equation}                                                        \label{LeavesRhoSSDiagonal}
  N(\Delta t)\rho_{\rm ss}[N(\Delta t)]^{\dag}
 =
                                       \left(\begin{array}{cc}
                                                 \Lambda_1&0\\
                                                 0& \Lambda_2
                                              \end{array}
                                        \right)\,.
\end{equation}
Using Theorem~1 we have
\begin{eqnarray}                                                                    \label{BothLambdas}
2e^{-2Z_1}\Lambda_1 &=& 
\Big(e^{-\frac{\gamma\Delta t}{2}a^{\dag}a}
         e^{Z_2^{+}a^{\dag}}
         e^{Z_3^{-}a}
 \Big)  |\alpha\rangle\langle\alpha|\Big(\cdots\Big)^{\dag}\,,\cr
2e^{-2Z_1}\Lambda_2 &=& 
\Big(e^{-\frac{\gamma\Delta t}{2}a^{\dag}a}
         e^{Z_2^{-}a^{\dag}}
         e^{Z_3^{+}a}
 \Big)  |\alpha^*\rangle\langle\alpha^*|\Big(\cdots\Big)^{\dag}\,,
\end{eqnarray}
where
\begin{eqnarray}
&&\hspace{-10mm}  
Z_2^{\pm}\equiv\frac{2E\pm ig}{\gamma}(e^{\gamma\Delta t/2}-1)\,,\cr
&&\hspace{-10mm}  
 Z_3^{\pm}\equiv\frac{2E\pm ig}{\gamma}(e^{-\gamma\Delta t/2}-1)\,.
\end{eqnarray}
In order to calculate $\Lambda_1$ we use the identity
$e^{\lambda a^{\dag}}|\alpha\rangle\langle\alpha |e^{\lambda^* a}
=e^{|\alpha+\lambda|^2-|\alpha|^2}
|\alpha+\lambda\rangle\langle\alpha+\lambda|$
which gives
\begin{equation}
2e^{-2Z_1}\Lambda_1=
|e^{Z_3^{-}\alpha}|e^{|\alpha+Z_2^{+}|^2-|\alpha|^2}
\Big( e^{-\frac{\gamma\Delta t}{2}a^{\dag}a}
 |\alpha+Z_2^{+}\rangle\langle\alpha+Z_2^{+}|
e^{-\frac{\gamma\Delta t}{2}a^{\dag}a}\Big)\,.
\end{equation}
Now, with the help of the identity $ e^{-\lambda a^\dag a}|\alpha\rangle\langle\alpha|e^{-\lambda a^{\dag}a}
=e^{|\alpha|^2(e^{-2\lambda}-1)}
|\alpha e^{-\lambda}\rangle\langle\alpha e^{-\lambda}|$ we have
\begin{equation}
  2e^{-2Z_1}\Lambda_1=
|e^{Z_3^{-}\alpha}|
e^{|\alpha+Z_2^{+}|^2 e^{-\gamma\Delta t} -|\alpha |^2  }
|(\alpha+Z_2^{+})e^{-\frac{\gamma \Delta t}{2}}\rangle
\langle(\alpha+Z_2^{+})e^{-\frac{\gamma \Delta t}{2}}|\,.
\end{equation}
Using the definition of $Z_2^{+}$ and the value of
$\alpha=(2E+ig)/\gamma$ we see that
\begin{equation}
(\alpha+Z_2^{+})e^{-\frac{\gamma \Delta t}{2}}=\alpha.
\end{equation}
Therefore
\begin{equation}
  2e^{-2Z_1}\Lambda_1=
|e^{Z_3^{-}\alpha}|
| \alpha\rangle\langle \alpha|\,.
\end{equation}
Repeating the same arguments for $\Lambda_2$ we have from
Eq.~(\ref{BothLambdas}):
\begin{equation}
   2e^{-2Z_1}\Lambda_2=
|e^{Z_3^{+}\alpha^*}|
| \alpha^*\rangle\langle \alpha^*|\,.
\end{equation}
Because $|e^{Z_3^{-}\alpha}|= |e^{Z_3^{+}\alpha^*}|$ we can now see that
\begin{equation}
   \left(\begin{array}{cc}
                                                 \Lambda_1&0\\
                                                 0& \Lambda_2
                                              \end{array}
                                        \right)
\propto \left(\begin{array}{cc}
                                                  | \alpha\rangle\langle \alpha|      &0\\
                                                 0& | \alpha^*\rangle\langle \alpha^*|
                                              \end{array}
                                        \right) = 2\rho_{\rm ss}\,.
\end{equation}
Together with Eq.~(\ref{LeavesRhoSSDiagonal}) this completes the proof.
$\Box$

\section{Conditional evolution starting from the steady state} \label{Sec:Cond}

In the previous section we have shown that, for a real value of $\beta$, a
homodyne measurement does not give any information about the system once
it has reached the steady state $\rho_{\rm ss}$. Although this fact was useful
in confirming that $\rho_{\rm ss}$ is indeed a steady state of the system,
such a measurement would be pointless in practice.

We therefore consider the case of purely imaginary $\beta$, for which the
homodyne measurement does provide information about the system. We write
$\beta$ in the form $\beta=i\beta_0$, where
$\beta_0$ is real. To find the conditional density matrix in this case, we go
back to Eqns.~(\ref{NewEquationForR}--\ref{aPlusF}) and obtain by direct
calculation:
\begin{equation}
  f(\sigma_y)[f(\sigma_y)]^{\dag}
         =\frac{4E^2+g^2}{\gamma^2}
         (e^{\gamma t_p/2}-1)^2 
           +\beta_0^2 e^{\gamma t_p}
           +(-1)^{k_p}\frac{2g\beta_0}{\gamma}
             (e^{\gamma t_p}-e^{\gamma t_p/2})\,,
\end{equation}                         
and
\begin{equation}
  {\rm Re}[f(\sigma_y)] {\rm Re}(\alpha) 
+{\rm Im}[f(\sigma_y)] {\rm Im}(\alpha)\sigma_y 
=
\frac{4E^2+g^2}{\gamma^2}(e^{\gamma t_p/2}-1)
     +(-1)^{k_p}\frac{g\beta_0}{\gamma} e^{\gamma t_p/2}\sigma_y\,.
\end{equation}
Therefore
\begin{eqnarray}
G(\rho_{\rm ss}, i\beta_0)
&=&
\int_0^{\Delta t}dt_m\int_0^{\Delta t}dt_{m-1}\cdots\int_0^{\Delta t}dt_1 \cr
&&\ \ \ \ \ \ \prod_{p=1}^{m}
\Big(
\frac{4E^2+g^2}{\gamma^2}+\beta_0^2
+(-1)^{k_p}\frac{2g \beta_0}{\gamma}
[\one+(\sigma_y-\one)e^{-\gamma t_p/2}]
 \Big)\rho_{\rm ss}\,.
\end{eqnarray}
Performing the integration we obtain
\begin{eqnarray}
  G(\rho_{\rm ss}, i\beta_0)
&=&\prod_{p=1}^{m}
\Big[\Delta t\cdot \Big(
\frac{4E^2+g^2}{\gamma^2}+\beta_0^2
+(-1)^{k_p}\frac{2g \beta_0}{\gamma}\Big)\cr
&&\ \ \ \ \ \ \ \ \ \ \ \ \ \ \ \ \ \ \ \ \ \ \  \ \ +
(-1)^{k_p} 4g \beta_0(\sigma_y-\one)
\frac{1- e^{-\gamma \Delta t/2}}{\gamma^2}
 \Big]\rho_{\rm ss}\;.
\end{eqnarray}
Because $\rho_{\rm ss}$ and $\sigma_y$ are both diagonal
in the basis $\{\ket{\pm}\}$ defined in Eq.~(\ref{basis}),
the conditional density matrix can be written in the form
\begin{equation}
\rho_{\rm c}(k_1,\dots,k_m;\Delta t)=  
                                   \left(\begin{array}{cc}
                       \lambda_1        |\alpha\rangle\langle\alpha|&0\\
                    0&  \lambda_2 |\alpha^*\rangle\langle\alpha^*|
                                              \end{array}
                                        \right)\,,
\end{equation}
where $\alpha=(2E+ig)/\gamma$. For the eigenvalues $\lambda_1$ and
$\lambda_2=1-\lambda_1$ we have the following simple formula:
\begin{equation}
\frac{\lambda_1}{\lambda_2}=
\prod_{p=1}^{m}\frac{b+(-1)^{k_p}}{b
   +(-1)^{k_p}(1-\frac{4}{\gamma\Delta t}
(1-e^{-\gamma\Delta t/2}))}\,,
\end{equation}
where $b\equiv (4E^2+g^2+\gamma^2\beta_0^2)/(2g\gamma\beta_0)$.  Similarly
simple expressions can be obtained for any complex reference field
$\beta$.

\section{Conclusion}   \label{Sec:Conclusion}

In this paper, we have given explicit formulas for the quantum state evolution
conditioned on a discrete homodyne photocount record for a typical
experimental setup in single-atom cavity QED.  These formulas have potential
applications for the real-time processing of experimental data.
The general methods developed here can be applied to a wide class of similar
systems. For example, it should be straightforward to generalize our results
to the case of heterodyne measurements.

\acknowledgements

We would like to thank Howard Wiseman for helpful comments on a
previous version of this manuscript. This work was supported by the EU
IST programme.

\section*{Appendix}

\noindent{\bf Theorem 1} 
{\it The operator 
\begin{equation}                                                                                       \label{defM}
M(t)=\exp\left[ig\frac{\sigma_y}{2}(a^\dag+a)t+E(a^{\dag}-a)t
                     -\frac{\gamma t}{2}a^{\dag}a \right]
\end{equation}
can be factorized as
\begin{equation}
M(t)= e^{Z_1}
            e^{-\frac{\gamma t}{2}a^{\dag}a}
            e^{Z_2 a^{\dag}}
            e^{Z_3 a}\;,
\end{equation}
where
\begin{eqnarray}
Z_1&=&\frac{4E^2+g^2}{\gamma^2}(1-e^{-\gamma t/2}-\gamma t/2) \cr
Z_2&=& \frac{2E+ig\sigma_y}{\gamma}(e^{\gamma t/2}-1)\cr
Z_3&=&\frac{2E-ig\sigma_y}{\gamma}(e^{-\gamma t/2}-1)
\end{eqnarray}}\\
{\bf Proof}\\
Because $a, a^\dag$, $a^\dag a$ and $\one$ span a Lie algebra, $M(t)$
can be factorized in a systematic way as follows. First we find
a function $x(t)$ such that
\begin{equation}                                                                               \label{FirstFactor}
M(t)=e^{x(t)a^{\dag}a}\tilde{M}(t)\;,
\end{equation}
where $\tilde{M}(t)$ is an exponential of a linear combination of
$a$ and $a^{\dag}$. We will then repeat the same procedure factorizing
$\tilde{M}$ which will conclude the prove of the theorem.

Equation (\ref{FirstFactor}) gives
\begin{equation}
\frac{d{M}}{dt}=\dot{x}a^{\dag}a\; e^{x a^{\dag}a}\tilde{M}
                  +e^{x a^{\dag}a}{d\tilde{M}\over dt}\;.
\end{equation}
On the other hand, Eq.~(\ref{defM}) gives
\begin{equation} 
\frac{d{M}}{dt}=[ig\frac{\sigma_y}{2}(a^\dag+a)+E(a^{\dag}-a)
                     -\frac{\gamma}{2} a^{\dag}a ]e^{x a^{\dag}a}\tilde{M}\;.
\end{equation}
Comparing this expression with the previous one we have
\begin{equation}                                                                                \label{dotTildeM}
{d\tilde{M}\over dt}= [\chi(x)
                     -(\dot{x}+\gamma/2) a^{\dag}a] \tilde{M}\;,
\end{equation}
where
\begin{equation}
\chi(x)=e^{-x a^{\dag}a}
                [ig\frac{\sigma_y}{2}(a^\dag+a)+E(a^{\dag}-a)]
               e^{x a^{\dag}a}\;.
\end{equation}
Using the identity $e^{-xa^{\dag} a}a e^{x a^{\dag}a}=ae^{x}$,
the above equation can be rewritten as
\begin{equation}
\chi(x)=(E+ig\frac{\sigma_y}{2})e^{-x} a^{\dag}
                 -(E-ig\frac{\sigma_y}{2})e^{x} a\;.
\end{equation}
Looking at Eq.~(\ref{dotTildeM}) we demand that
\begin{equation} \label{ForX}
\dot{x}+\gamma/2=0\;,
\end{equation}
thereby making $d\tilde{M}/dt$ independent of $a^{\dag}a$.
From Eq.~(\ref{defM}) we see that $M(0)=\one$ and therefore,
we choose, in accordance with Eq.~(\ref{FirstFactor}), that
\begin{equation} 
x(0)=0 {\mbox{\rm \ \ and\ \ }} \tilde{M}(0)=\one\;.
\end{equation}
With these conditions equation (\ref{ForX})
can be integrated to give, according to Eqns.~(\ref{FirstFactor})
and (\ref{dotTildeM}),
\begin{equation}
M(t)=e^{-\frac{\gamma t}{2}a^{\dag}a}\tilde{M}(t)\;,
\end{equation}
where
\begin{equation}                                                                                      \label{tildeM}
{d\tilde{M}\over dt}=
[(E+ig\frac{\sigma_y}{2})e^{\gamma t/2}a^{\dag}
-(E-ig\frac{\sigma_y}{2})e^{-\gamma t/2}a
]e^{ya^{\dag}}\tilde{M}'\;.
\end{equation} 
The proof of the theorem will be completed if we repeat the same
procedure for factorizing $\tilde{M}$. As before we introduce a function
$y(t)$ such that
\begin{equation}                                                                         \label{SecondFactor}
\tilde{M}(t)=e^{y(t)a^{\dag}}\tilde{M}'(t)\,.
\end{equation}
We therefore have
\begin{equation}
{d\tilde{M}\over dt}
=\dot{y}a^{\dag}e^{ya^{\dag}}\tilde{M}'
                              +e^{y a^{\dag}}{d\tilde{M}'\over dt}\;.
\end{equation}
Combined with Eq.~(\ref{tildeM}) this gives
\begin{equation}
{d\tilde{M}'\over dt}=
[(E+ig\frac{\sigma_y}{2})e^{\gamma t/2}a^{\dag}
-(E-ig\frac{\sigma_y}{2})e^{-\gamma t/2} e^{-ya^{\dag}}ae^{ya^{\dag}}
-\dot{y}a^{\dag}
]\tilde{M}'\,.
\end{equation}
Using the identity $e^{-ya^{\dag}}ae^{ya^{\dag}}=a+y$
we rewrite the above expression as
\begin{equation}                                                                    \label{dotTildeMprime}
{d\tilde{M}'\over dt}=
\left([(E+ig\frac{\sigma_y}{2})e^{\gamma t/2}-\dot{y}] a^{\dag}
-(E-ig\frac{\sigma_y}{2})e^{-\gamma t/2} (a+y)
\right)\tilde{M}'\,.
\end{equation}
We eliminate $a^{\dag}$ from this expression by setting
\begin{equation}                                                                                          \label{dotY}
\dot{y}=(E+ig\frac{\sigma_y}{2})e^{\gamma t/2}\,.
\end{equation}
Equation (\ref{SecondFactor}) suggests the boundary conditions 
\begin{equation}
y(0)=0 {\mbox{\rm \ \ and\ \ }} \tilde{M}'(0)=\one\;.
\end{equation}
Performing integration in (\ref{dotTildeMprime}) and in (\ref{dotY})
using these boundary conditions and the fact that $\sigma_y^2=\one$
we have according to (\ref{SecondFactor})
\begin{equation}
\tilde{M}(t)=\exp[\frac{2E+ig\sigma_y}{\gamma}(e^{\gamma t/2}-1) 
                                    a^{\dag}]\tilde{M}'(t)\,,
\end{equation} 
where
\begin{equation}
\tilde{M}'(t)=\exp[\frac{4E^2+g^2}{\gamma^2}
                                    (1-e^{-\gamma t/2}-\gamma t/2)]
                         \exp[-\frac{2E-ig\sigma_y}{\gamma}(1-e^{-\gamma t/2})a]\,.
\end{equation}
This completes the proof of the theorem.
$\Box$ \\

\noindent{\bf Corollary}
\begin{equation}
e^{-iH_0t}=e^{-g^2t^2/8}e^{igt\sigma_y a^\dag/2}
                                                  e^{igt\sigma_y a/2}\,.
\end{equation}
{\bf Proof}\\
This can be established easily by repeating the arguments of Theorem 1 for
$E=0$ and $\gamma=0$. $\Box$ \\

\noindent{\bf Theorem 2}
{\it Using the definition
\begin{equation}                                                                                               \label{fk}
f_k\equiv\sqrt{\gamma/2}\,e^{i\pi(k-1)/2}
\end{equation}
and the notation of Theorem 1, we have
\begin{equation}                                                                                          \label{CkM}
C_kM(t)=M(t) f_k  [e^{-\gamma t/2}a
                                    +\frac{1-e^{-\gamma t/2}}{\gamma}
                                       (2E+ig\sigma_y)+(-1)^k\beta]\,.
\end{equation}}\\
{\bf Proof}\\
By definition [Eqns. (\ref{Ck}) and (\ref{fk})] and using
Theorem~1 we have
\begin{equation}
C_kM(t)=f_ke^{Z_1}[a+(-1)^k\beta]
            e^{-\frac{\gamma t}{2}a^{\dag}a}
            e^{Z_2 a^{\dag}}
            e^{Z_3 a}\;,
\end{equation}
where $Z_1$, $Z_2$ and $Z_3$ are specified in the statement of
Theorem~1. Using subsequently the identities
$e^{-xa^{\dag} a}a e^{x a^{\dag}a}=ae^{x}$
and then $e^{-ya^{\dag}}ae^{ya^{\dag}}=a+y$ we have
\begin{eqnarray}
C_kM(t)&=&f_ke^{Z_1} 
            e^{-\frac{\gamma t}{2}a^{\dag}a}
                 [e^{-\gamma t/2}a+(-1)^k\beta]
            e^{Z_2 a^{\dag}}
            e^{Z_3 a}\cr
&=&f_ke^{Z_1} 
            e^{-\frac{\gamma t}{2}a^{\dag}a}
            e^{Z_2 a^{\dag}}
                 [e^{-\gamma t/2}(a+Z_2)+(-1)^k\beta]
            e^{Z_3 a}\cr
&=&f_kM(t) [e^{-\gamma t/2}(a+Z_2)+(-1)^k\beta]\,.
\end{eqnarray}
Putting the value of $Z_2$ from Theorem~1 we have
Eq.~(\ref{CkM}) as required.
$\Box$\\

%\bibliographystyle{prsty}
%\bibliography{andrei.bib}

\end{document}